\documentclass{tMPH2e}

\begin{document}

\doi{10.1080/00268976.2013.824126}
 \issn{1362–3028}
\issnp{0026–8976}
\jvol{00}
\jnum{00} \jyear{2013} %

\title{Hamiltonian replica-exchange in GROMACS:\break a flexible implementation}

\author{Giovanni Bussi\thanks{$^\ast$Corresponding author. Email: bussi@sissa.it
\vspace{6pt}} \\\vspace{6pt}  
{\em{Scuola Internazionale Superiore di Studi Avanzati (SISSA),\break via Bonomea
265, 34136 Trieste, Italy}}\\\vspace{6pt}\received{April 30, 2013} }

\maketitle

\begin{abstract}
A simple and general implementation of Hamiltonian
replica exchange for the popular molecular-dynamics software GROMACS
is presented.
In this implementation, arbitrarily different Hamiltonians can be used
for the different replicas without incurring in any significant
performance penalty.
The implementation was validated on a simple toy model - alanine
dipeptide in water - and applied to study the rearrangement of an RNA tetraloop,
where it was used to compare recently proposed force-field corrections.
\bigskip

\begin{keywords}
  Hamiltonian replica exchange; solute tempering; RNA tetraloop
\end{keywords}\bigskip

\end{abstract}

\section{Introduction}

Molecular dynamics (MD) is a powerful tool which can be used to simulate
the time evolution of molecular systems with great accuracy. However,
its application to realistic problems suffers from the so-called time-scale
issue. Indeed, whereas typical trajectories simulated with empirical
force fields are nowadays on the order of 1 $\mu$s long, interesting
events such as phase transitions, chemical reactions and conformational
changes often need much longer time scales. In spite of the development
of fast and scalable MD software \cite[see, e.g.,][]{hess+08jctc} and
\emph{ad hoc} hardware \cite[see, e.g.,][]{shaw+10science},
many interesting problems
can be expected to remain
unaffordable with direct MD simulations for several decades.
This issue has pushed the development of many techniques that allow
to effectively accelerate MD so as to be able to study relevant problems
with relatively low computational effort. A class of methods is based
on the idea of choosing \emph{a priori} a small set of collective
variables which are then biased during the simulation (e.g. umbrella
sampling \cite{torri-valle77jcp} and metadynamics \cite{laio-parr02pnas}).
A common problem with these techniques is that their efficiency and
accuracy is determined by the choice of the collective variables,
which is often a difficult task. Another class of methods is based
on the idea of raising temperature so as to accelerate sampling (e.g.
simulated tempering \cite{mari-pari92el} and parallel tempering \cite{hans97cpl,sugi-okam99cpl}).
In parallel tempering several replicas of the same system are simulated
at different temperature, and coordinates are exchanged from time
to time with a Monte Carlo procedure. A very well known issue of parallel
tempering is that the number of replicas required to span a preassigned
temperature range grows with the system size. More generally, all
these temperature-based methods suffer from the fact that the entire
system under investigation is accelerated. Thus, while they do not
require any \emph{a priori} knowledge of the investigated events and
can be often used blindly, they also do not allow the user to embed
any knowledge about the problem in the simulation setup. Hamiltonian
replica-exchange (HREX) methods \cite{sugi-okam00cpl}, where different
replicas evolve according to different Hamiltonians, are in a sort
of intermediate position among the two mentioned classes of methods,
and thus provide an interesting compromise among them. On the one
hand, they are simpler to use when compared with collective-variable
based methods. Indeed, dependence of the results on the choice of
the modified Hamiltonian is smaller than dependence of, say, umbrella
sampling efficiency on the choice of the collective variables. On the
other hand, they are more efficient than parallel tempering because
the number of required replicas is typically much less. A wealth of
recipes for HREX has been proposed in the last years
\cite[see, among others,][]{sugi-okam00cpl,fuku+02jcp,liu+05pnas,affe+06jctc,faje+08jctc,xu+08jctc,zach08jctc,vree+09jpcb,itoh+10jcp,lee+10jctc,meng-roit10jctc,tera+11jcc,wang+11jpcb,zhang2012folding}.

We here focus on one of the most successful among these
recipes, namely replica-exchange solute tempering in its REST2 variant
\cite{wang+11jpcb}, and discuss an implementation of this method
in the popular MD software GROMACS \cite{hess+08jctc}. We also discuss
a possible extension of REST2 where only a part of the solute is modulated.
The implementation is validated on alanine dipeptide in water and
applied to study the stability of an RNA tetraloop, comparing two
recently developed force fields \cite{pere+07bj,zgar+11jctc}.

\section{Methods}

\subsection{Hamiltonian replica exchange}

We consider a system with coordinates $r$ and subject to a potential
energy $U(r)$. We assume that the potential is built as a sum of
few-body terms as it is conventionally done for the atomistic modeling
of biomolecules \cite{corn+95jacs,chea+99jbsd}, although the method can
be easily generalized to other force fields. The system is assumed
to be coupled with a thermal bath at temperature $T$ so that the
probability of exploring a configuration is $P(r)\propto e^{-\frac{U(r)}{k_{B}T}}$
where $k_{B}$ is the Boltzmann constant. Replica exchange methods
are generally based on the idea of sampling one ``cold'' replica,
from which the unbiased statistics can be extracted, plus a number
of ``hot'' replicas, whose only purpose is that of accelerating sampling.
The ``hottest'' replica should explore the space fast enough to overcome
barriers for the process under investigation, whereas the intermediate
replicas are necessarily introduced to bring the system smoothly from
the ``hottest'' ensemble to the ``coldest'' ensemble. Indeed, the number
of needed replicas is actually related to the difference between the
hottest and the coldest ensembles. In plain parallel tempering ``hot''
and ``cold'' refers to physical temperature as controlled by a thermostat,
whereas in the general Hamiltonian replica exchange ``hot'' replicas
can be biased in an arbitrary manner so as to accelerate sampling.
It should be noted that transition rates for processes which are hindered
by entropic barriers are not necessarily expected to increase with
temperature, so that the efficiency of parallel tempering in those
cases could be lower\emph{ }\cite{nyme+08jctc}.

In the most general formulation, each replica is simulated at a different
temperature and using a different Hamiltonian. Calling $r_{i}$ the
coordinate of the $i$-th replica and $N$ the number of replicas,
the resulting product ensemble is 
\begin{eqnarray*}
P(r_{1})\times\dots\times P(r_{N}) & \propto & e^{-\frac{U_{1}(r_{1})}{k_{B}T_{1}}-\dots-\frac{U_{N}(r_{N})}{k_{B}T_{N}}}
\end{eqnarray*}
If $U_{1}=U_{2}=\dots=U_{N}$, plain parallel tempering is recovered.
Since the ensemble probability only depends on $U/(k_{B}T)$, a double
temperature is completely equivalent to a halved energy. The advantage
of scaling the potential energy instead of the temperature is related
to the fact that the energy is an extensive property, whereas the
temperature is an intensive one. One can thus selectively choose a
portion of the system and specific parts of the Hamiltonian to be
``heated.'' Still there is some arbitrariness in the scaling of the
coupling terms. In our approach, we split the system in two regions
$\mathcal{H}$ (hot) and $\mathcal{C}$ (cold) so that each atom is
statically assigned to either the $\mathcal{H}$ or the $\mathcal{C}$ region, and defined a parametrized
Hamiltonian which depends on $\lambda$ as follows: 
\begin{itemize}
\item The charge of atoms in the $\mathcal{H}$ region is scaled by a factor
$\sqrt{\lambda}$. 
\item The $\epsilon$ (Lennard-Jones parameter) of atoms in the $\mathcal{H}$
region is scaled by a factor $\lambda$. 
\item The proper dihedral potentials for which the first \emph{and} fourth
atoms are in the $\mathcal{H}$ region is scaled by a factor $\lambda$. 
\item The proper dihedral potentials for which either the first \emph{or}
the fourth atom is in the $\mathcal{H}$ region is scaled by a factor
$\sqrt{\lambda}$. 
\end{itemize}
With this choice, only force-field terms contributing to energy barriers,
i.e. electrostatic, Lennard-Jones and proper dihedrals, are scaled
in such a manner that:
\begin{itemize}
\item Interactions inside the the $\mathcal{H}$ region are kept at an effective
temperature $T/\lambda$.
\item Interactions between the $\mathcal{H}$ and the $\mathcal{C}$ regions are
kept at an effective intermediate temperature $T/\sqrt{\lambda}$.
\item All interactions inside the $\mathcal{C}$ region are kept at temperature
$T$.
\end{itemize}
We underline that the effective temperature is not induced by a thermostat and that
the simulations as well as the exchanges among replicas are performed
at thermodynamics equilibrium.
The scaling parameter $\lambda$ can be chosen to be any real number between 1, for the reference,
unmodified system, and 0. The latter value corresponds to no interaction
in the $\mathcal{H}$ region or, equivalently, to infinite temperature.
Albeit the code allows choosing $\lambda=0$ (infinite temperature),
this is usually not an optimal choice because it would lead to a very
low acceptance rate.
We also notice that if the $\mathcal{H}$
region has a net charge, the ``hot'' replicas will also have a total
charge which is different from that of the unbiased replica. This
is not a problem because in periodic calculations based on Ewald-like
methods \cite{dard+93jcp} a neutralizing background is implicitly
added. As a final observation, our choice for the treatment of scaling parameters
for dihedrals leads to a consistent scaling of dihedral potentials
and corresponding 1-4 interactions.

When used for the entire solute, our implementation exactly reproduces
REST2 \cite{wang+11jpcb}. Moreover, it was constructed in such a manner
to be used in a partial tempering scheme, where only a portion of
the solute is heated. It is straightforward to extend the formulation
so as to simulate replicas at different pressures \cite{okab+01cpl}.

\subsection{Implementation details}

We implemented our replica exchange methodology in GROMACS 4.6.1 \cite{hess+08jctc} patched with the PLUMED plugin \cite{bono+09cpc}, version 2.0b0.
The combination of GROMACS and PLUMED was used to allow
HREX and enhanced sampling methods
based on biasing \emph{a priori} chosen collective variables to be
used simultaneously.
To increase the
flexibility of the method, we coded it in such a manner that independent
topology files can be used for different replicas. In principle,
different PLUMED input files could also be used, thus adding different
bias potentials or restraints on the different replicas, albeit we
did not exploit this feature here.

\begin{figure}
\center
\includegraphics[width=0.5\columnwidth]{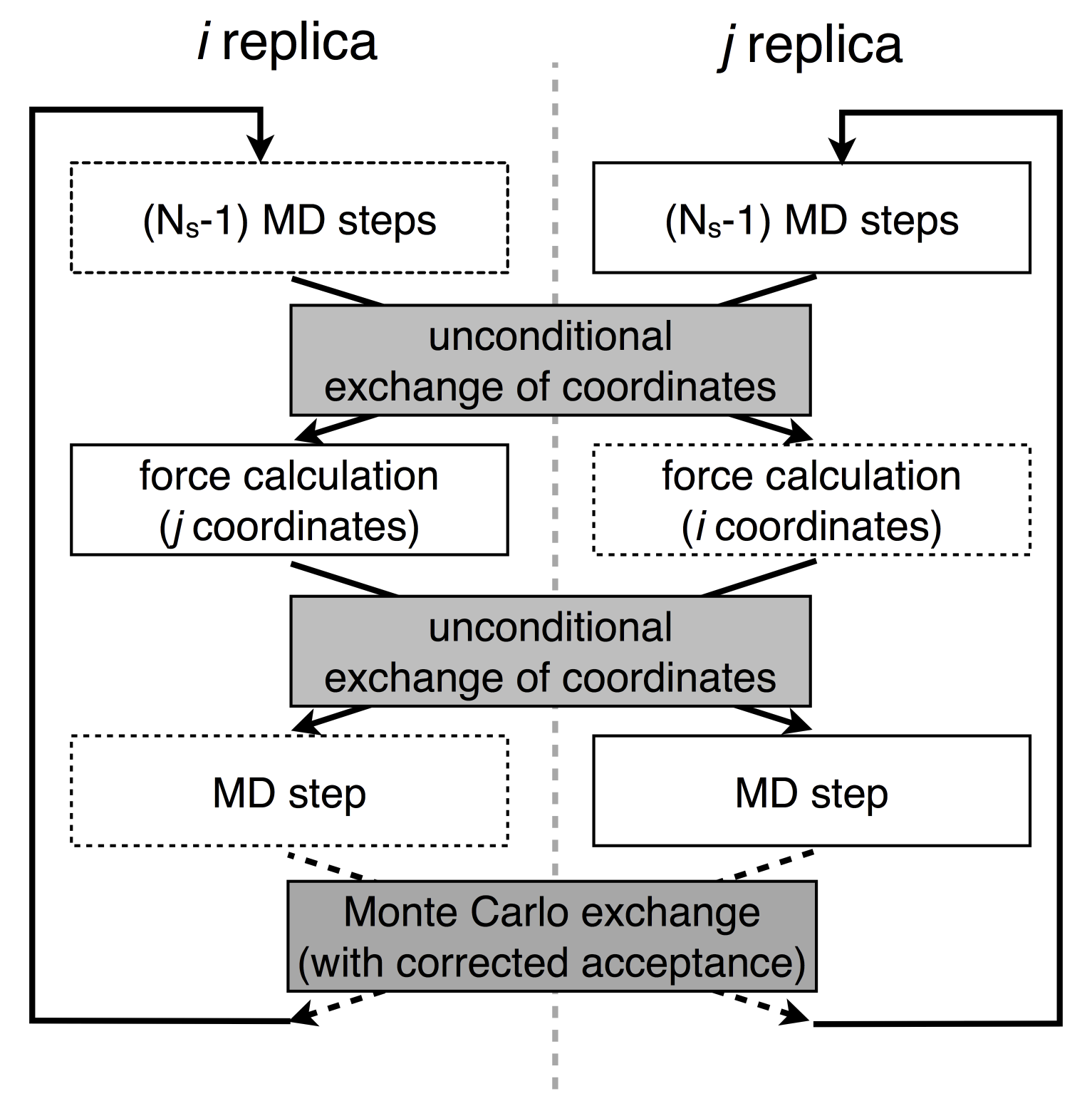}

\caption{Flowchart of our Hamiltonian replica exchange implementation. After
having performed $N_{s}-1$ molecular dynamics steps, a
coordinate swap is performed. Then, the energy is recomputed and
coordinates are swapped again.
At this point a further MD step is done
and a real exchange is attempted with a corrected Monte Carlo acceptance
[Equation \eqref{eq:acceptance}].\label{fig:flowchart}}
\end{figure}

The flow of the modified replica exchange is depicted in Figure \ref{fig:flowchart}.
 At the beginning of a timestep where an exchange is required an
exchange is unconditionally performed and the
total energy
is computed using the local force field
for the coordinates obtained from 
another processor.
This energy is stored for later usage, and original (unswapped) coordinates
are restored with an extra unconditional exchange. At the end of the
MD step, when the actual exchange is attempted, the previously stored
energy is used to accept/reject the exchange. The acceptance is then
computed in the most general manner, which allows replicas with different
bias potential, Hamiltonian and temperature
\begin{equation}
\alpha=\text{min}\left(1,e^{\frac{-\bar{U}_{i}\left(r_{j}\right)+\bar{U}_{i}\left(r_{i}\right)}{k_{B}T_{i}}+\frac{-\bar{U}_{j}\left(r_{i}\right)+\bar{U}_{j}\left(r_{j}\right)}{k_{B}T_{j}}}\right)\label{eq:acceptance}
\end{equation}
where $\bar{U}$ is defined as the sum of the force-field potential
and possibly additional potentials as computed by PLUMED.
Force-field parameters for the ``hot'' replicas are edited using simple
scripts. In spite of the two extra swaps required at each attempted
exchange, the overhead is rather low. Its exact value depends on the
attempt frequency for replica exchange, and in our experience never
exceeded 10\%.

We observe that our implementation differs from the one proposed in
Ref. \cite{tera+11jcc}, where the free-energy perturbation method
already available in GROMACS has been exploited. Because of the way interactions
for $0<\lambda<1$ are treated in GROMACS, strictly speaking it is
not possible to set up a simulation following REST2 prescriptions
using free-energy perturbation. Moreover, calculation of non-bonded
interactions in the free-energy perturbation are slower in
GROMACS, and can introduce significant overhead even in the plain
MD which is performed between exchanges. On the other hand, the overhead
of our implementation is limited to the exchange step. Since the stride
between exchanges is typically on the order of at least 100 steps,
this overhead is negligible.

\section{Applications}

\subsection{Solute tempering: alanine dipeptide\label{sub:alanine-dipeptide}}

As a first test case we focused on alanine dipeptide, a standard benchmark
for enhanced-sampling methods. The low-energy conformations of this
system can be described using the two dihedral angles of the Ramachandran
plot, $\phi$ and $\psi$. Transitions between conformations $C_{7\mbox{eq}}$
$(\phi=-80^{\circ},\psi=75^{\circ})$ and $C_{7\mbox{ax}}$
$(\phi=75^{\circ},\psi=-75^{\circ})$ are hindered by large
free-energy barriers.  An alanine dipeptide molecule modeled with
Amber99sb force-field \cite{lind+10pro} was solvated in a box containing
approximately 700 TIP3P water molecules \cite{jorg+83jcp}. All bonds
were kept rigid \cite{miya-koll92jcc,hess+97jcc}, and equations of
motion were integrated using a timestep of 2fs. Long-range electrostatics
was treated using particle-mesh Ewald \cite{essm+95jcp}, and temperature
was controlled by stochastic velocity rescaling \cite{buss+07jcp}.

\begin{figure}
\center
\includegraphics[clip,width=1\columnwidth]{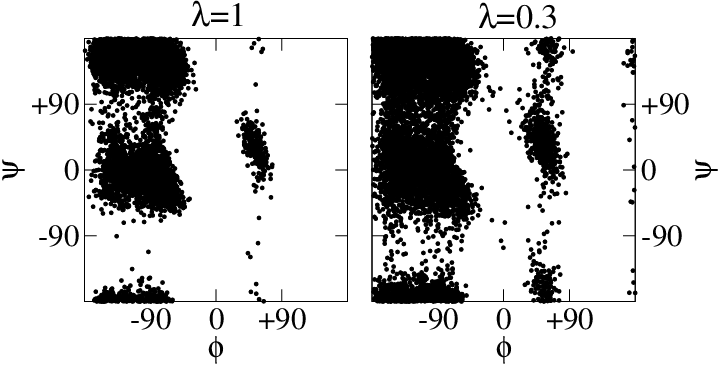}

\caption{\label{fig:ramachandran}Conformational space explored for alanine
dipeptide by first ($\lambda=1$, left) and last ($\lambda=0.3$,
right) replica. It can be seen that the conformational space explored
by last replica is larger.}
\end{figure}

\begin{figure}
\center
\includegraphics[clip,width=0.9\columnwidth]{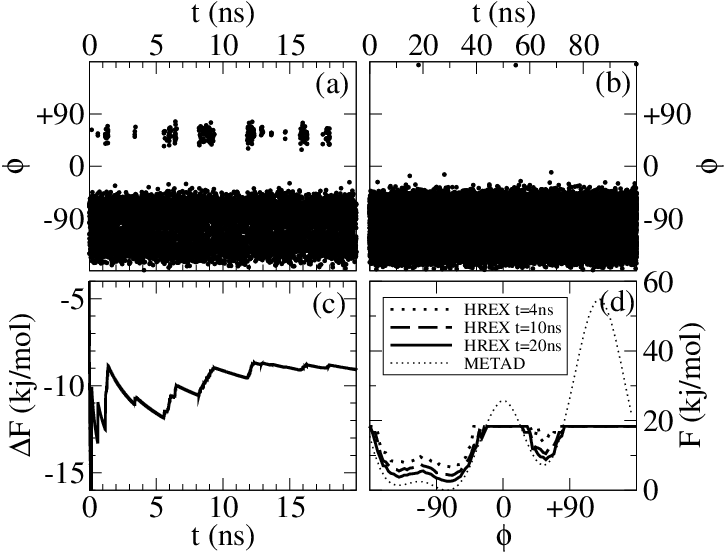}

\caption{\label{fig:convergence-ala}Convergence of Hamiltonian replica exchange
(HREX) for alanine dipeptide. $\phi$ angle for (a) replica at $\lambda=1$
and (b) for a longer, serial simulation.
(c) Estimate of the free-energy difference between $C_{7\mbox{eq}}$
and $C_{7\mbox{ax}}$ as a function of the simulated time per replica,
obtained from analyzing the replica at $\lambda=1$. (d) Free-energy
landscape as a function of dihedral angle $\phi$, as obtained from
HREX, compared with a reference metadynamics calculation. Results
for HREX are shown for different simulation lengths (simulated time
per replica equal to 4, 10 and 20 ns, as indicated), whereas metadynamics
profile has been obtained from a single 10 ns simulation.}
\end{figure}

We performed a REST2 \cite{wang+11jpcb} simulation using 5 replicas
with values of $\lambda$ ranging from 1 to 0.3 following a geometric
distribution. This choice lead to an acceptance rate ranging from
35\% to 50\%. Exchanges are attempted every 100 steps. In Figure \ref{fig:ramachandran}
the distributions of $\psi$ and $\phi$ angles explored by the first
and last replica are shown. It can be seen how the change in the Hamiltonian
effectively raises the temperature of the molecule, thus decreasing
the impact of free-energy barriers. The time series of the $\phi$
dihedral angle in the replica with $\lambda=1$ is shown in Figure
\ref{fig:convergence-ala}, together with a much longer single-replica
simulation. At the price of a factor 5 in the computational cost,
the HREX simulation sampled the phase space much faster. Since several
transitions between $C_{7\mbox{eq}}$ and $C_{7\mbox{ax}}$ are observed,
we could compute the relative stability of the two metastable minima,
which converged quickly (Figure \ref{fig:convergence-ala}c). The
free-energy profile as a function of the $\phi$ dihedral angle was
also computed and compared with a reference free-energy landscape
obtained using well-tempered metadynamics \cite{bard+08prl} (well-tempered
factor $\Delta T$=2100K, initial deposition rate $\omega$=6.25 kj/mol/ps,
Gaussian width $\sigma=20^{\circ}$, simulation length 10ns). Profiles
obtained at different stages of the HREX simulation and reference
metadynamics results are shown in Figure \ref{fig:convergence-ala}d.
For this simple system, metadynamics has the advantage of providing
good statistics also on the free-energy barriers. However, HREX is
capable of reproducing the correct free-energy difference between
the two minima and the correct shape of the two free-energy wells
using a minimal information about the simulated system. This can be
an advantage in cases where choosing collective variables is more difficult,
such as the one discussed below.

\subsection{Partial tempering: RNA tetraloop}

The second application is the structural characterization of a UUCG
RNA tetraloop. UUCG tetraloops and small RNA hairpins have been characterized
\emph{in vitro} and \emph{in silico }by several groups \cite{ma+06jacs,garc-pasc08jacs,zuo+10jpcb,bana+10jctc,kuhr+13jctc}.
Atomistic molecular simulations of tetraloop folding are difficult
because of slow sampling and of the well-known inaccuracies of classical
force fields for RNA \cite{bana+10jctc}. In a recent paper, Kurova
\emph{et al.}~\cite{kuhr+13jctc}
have shown the results of a long parallel-tempering simulation
of a UUCG tetraloop. In their work, the full hairpin
is initialized in a straight conformation, so as to blindly predict
its folded structure and stability. Our investigation was instead
limited to the exploration of the conformational space available for
the tetraloop, without studying the full hairpin formation.

\begin{figure}
\center
\includegraphics[clip,width=0.8\columnwidth]{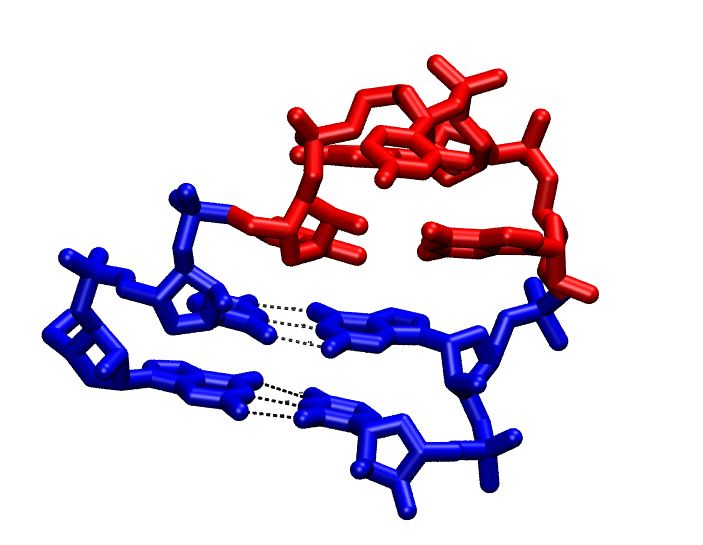}
\caption{\label{fig:RNA}
Representation of the RNA tetraloop, hydrogen atoms not shown. Atoms
in the ``hot'' region (tetraloop) are shown in red. Atoms
in the ``cold'' region (stem) are shown in blue.
Restrained Watson-Crick hydrogen bonds are also marked.
Graphics made with VMD \cite{humphrey1996vmd}.
}
\end{figure}

We started from an experimental structure (residue 31-38 of PDB 1F7Y
\cite{enni+00jmb}, sequence GCUUCGGC) solvated in a box containing
approximately 4600 water molecules and added 14 Na$^{+}$ and 7 Cl$^{-}$
atoms. All other simulation details were chosen as in Subsection~\ref{sub:alanine-dipeptide}.
After equilibration, we restrained the 6 Watson-Crick hydrogen bonds
of the stem (enforced distance 3$\mbox{\AA}$, stiffness 25 kj/mol/$\mbox{\AA}^{2}$;
see Figure~\ref{fig:RNA})
so as to suppress fraying of the first base pair and avoid severe
unfolding of the hairpin. We selected as a ``hot'' region the 4 nucleotides
corresponding to the tetraloop (see Figure~\ref{fig:RNA}), leaving the Hamiltonian for the stem
unbiased in all the replicas. We simulated 16 replicas using values
of $\lambda$ ranging from 1 to 0.3 with a geometric distribution,
leading to an acceptance rate which is between 30\% and 50\%. This
protocol allowed us to accelerate the sampling of different conformations
of the tetraloop, without perturbing too much the stem. Simulations were performed
using two recently developed force fields, both based on Amber99 force
field \cite{chea+99jbsd}: parmbsc0 force field \cite{pere+07bj}
(from now on, bsc0) and ff99bsc0$_{\chi\mbox{OL3}}$ force
field \cite{zgar+11jctc} (from now on, bsc0-OL).

\begin{figure}
\center
\includegraphics[clip,width=0.9\columnwidth]{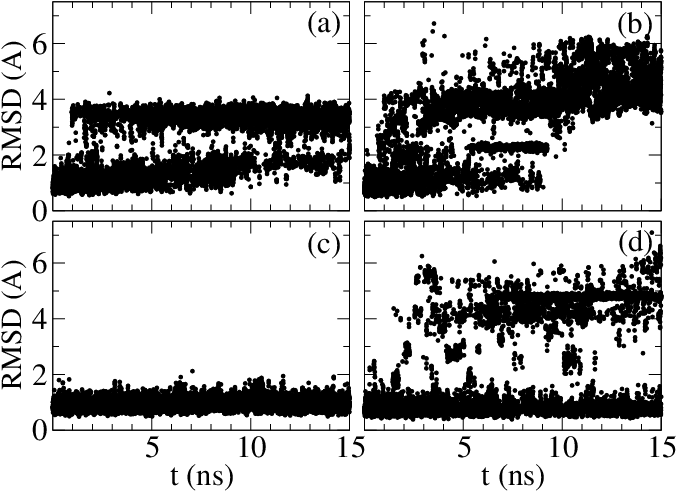}

\caption{\label{fig:RMSD-RNA}Root-mean-square deviation of hairpin stem (bases
1, 2, 7 and 8) and loop (bases 3, 4, 5, 6) from the experimental structure,
as obtained from the unbiased replica ($\lambda=1$) as a function
of simulation time per replica. Simulations were performed
using bsc0 force field
{[}stem (a) and loop (b){]} and bsc0-OL force field {[}stem
(c) and loop (d){]}. It can be seen that the latter force field better
stabilizes
the native structure for the stem. Loop stability is also improved
with bsc0-OL, but in this case also non-native, high RMSD, structures
are sampled.}
\end{figure}

In Figure \ref{fig:RMSD-RNA} the root-mean-square deviation (RMSD)
of stem and loop from the reference experimental structure is shown.
The simulation performed using the bsc0 force-field quickly
interconverted into an artificial ``ladder-like'' structure \cite{mlyn+10jpcb}
for which the RMSD of the stem from the experimental structure is
$\approx$4$\textrm{\AA}$.
This is a known problem of the bsc0 force
field, and has been already detected by means of long MD simulations
\cite[see][and references therein]{bana+10jctc}.
Notably, with Hamiltonian replica exchange this happened in a very short
time ($\approx1$ns per replica).
The coexistence
of a correct (low RMSD) and artificial (high RMSD) structure in the
Figure \ref{fig:RMSD-RNA}a is due to the fact that the only the replica
at $\lambda=1$ is shown. More precisely, some trajectories switched
to the ``ladder-like'' structure, and other ones did not, resulting
in a mixed ensemble for the $\lambda=1$ replica. The native loop
structure was even less stable: After approximately 10 ns per replica
the native structure was destroyed in all replicas and completely
disappeared from
the explored ensemble (Figure \ref{fig:RMSD-RNA}b). On the other
hand, the simulation performed using the bsc0-OL force field
behaved in a qualitatively better way. The stem was very stable on
the same timescale (Figure \ref{fig:RMSD-RNA}c), and, even if spurious
structures were appearing in the loop, the native structure was still
populated after 15 ns per replica (Figure \ref{fig:RMSD-RNA}d). This indicates
that the actually explored ensemble and the experimental one are reasonably
overlapping.

These results show that Hamiltonian replica exchange, especially
in variants were only a portion of a larger molecules is biased, can
be very effective in accelerating conformational sampling.
In particular, we were able to detect the known problems
of the bsc0 force field in a short computational time. A deeper
investigation of the force-field dependence of the conformational
space available for an RNA tetraloop will be the
subject of further investigations.

\section{Conclusions}

In conclusion, a flexible implementation of Hamiltonian
replica exchange for GROMACS was discussed.
This implementation can be used to combine
replicas at different temperature, pressure, and using different force-fields.
It was validated on the simple case of alanine dipeptide in water,
where results obtained with a reference
well-tempered metadynamics calculation
were correctly reproduced. Then, Hamiltonian
replica exchange was used to extensively sample the available conformations
in an RNA tetraloop, comparing two different force fields. Our software
is available upon request and will be distributed together with the
next release of PLUMED.

\section*{Acknowledgments}

The research leading to these results has received
funding from the European Research Council under the European
Union's Seventh Framework Programme (FP/2007-2013) /
ERC Grant Agreement n. 306662, S-RNA-S.
Simulations were performed partly on local machines and partly at
the CINECA supercomputing center. The ISCRA grant HP10B194XL and the
IIT Platform ``Computation'' are acknowledged for the availability
of high performance computing resources. Sandro Bottaro and Maria
Darvas are acknowledged for providing a port of the bsc0-OL force
field for GROMACS.
Sandro Bottaro is also acknowledged for carefully reading the manuscript.
Luca Bellucci and Ben Cossins are acknowledged
for testing preliminary versions of this replica-exchange implementation.

\label{lastpage}

\end{document}